\documentstyle[12pt]{article}
\def\d{\delta}
\def\e{\epsilon}

\def\p{\psi}

\def\l{\lambda}

\def\t{\theta}

\def\bt{\bar{\t}}

\def\bp{\bar{\p}}

\def\be{\begin{equation}}
\def\ee{\end{equation}}
\def\arr{\begin{array}{rll}}
\def\ea{\end{array}}
\def\bea{\begin{eqnarray}}
\def\eea{\end{eqnarray}}

\newcommand{\q}[1]{(\ref{#1})}
\newcommand{\nn}{\nonumber \\}

\begin{document}
%\large
\begin{titlepage}

\vskip 4.0cm
\renewcommand{\thefootnote}{\fnsymbol{footnote}}
\begin{center}

{\Large\bf Quantum mechanics of superparticle with 1/4
supersymmetry breaking}\\

\bigskip

\vskip 2cm

S. Bellucci~$^{a,}$\footnote{bellucci@lnf.infn.it},
A. Galajinsky~$^{a,}$\footnote{On leave from Tomsk Polytechnical
University, Tomsk, Russia\\
\phantom{XX}
agalajin@lnf.infn.it, galajin@mph.phtd.tpu.edu.ru},
E. Ivanov~$^{b,}$\footnote{eivanov@thsun1.jinr.ru} and
S. Krivonos~$^{b,}$\footnote{krivonos@thsun1.jinr.ru}

\vskip 0.5cm

$^{a)}$ {\it INFN--Laboratori Nazionali di Frascati, C.P. 13,
00044 Frascati, Italy}

\vskip 0.2cm

$^{b)}$ {\it Bogoliubov Laboratory of Theoretical Physics, JINR,\\
141 980, Dubna, Moscow Region, Russian Federation}

\end{center}
\vskip 1cm
\renewcommand{\thefootnote}{\arabic{footnote}}
\setcounter{footnote}0
\begin{abstract}
We study quantum mechanics of a massive superparticle in $d=4$
which preserves 1/4 of the target space supersymmetry with eight
supercharges, and so corresponds to the partial breaking
$N=8 \rightarrow N=2$. Its worldline action contains a Wess-Zumino term,
explicitly breaks $d=4$ Lorentz symmetry and exhibits
one complex fermionic $\kappa$-symmetry.
We perform the Hamiltonian analysis of the model and quantize it in two different
ways, with gauge-fixed $\kappa $-symmetry and in the Gupta-Bleuler
formalism. Both approaches give rise to the same supermultiplet structure
of the space of states. It contains three irreducible $N=2$ multiplets
with the total number of (4 +4) complex on-shell components. These states
prove to be in one-to-one correspondence with the de Rham complex of
$p$-forms on a three-dimensional subspace of the target $x$-manifold.
We analyze the vacuum structure of the model and find that
the non-trivial vacua are given by the exact harmonic one- and two-forms.
Despite the explicit breaking of $d=4$ Lorentz symmetry in the
fermionic sector, the $d=4$ mass-shell condition is still valid
in the model.
\end{abstract}

\vspace{0.5cm}

PACS: 04.60.Ds; 11.30.Pb

{\it Keywords}: Partial breaking of supersymmetry; Intersecting
branes; Superparticle

\end{titlepage}

%\noindent

\section{Introduction}
%\vskip 0.3cm

Nowadays, partial breaking of global supersymmetry (PBGS) \cite{bw,p}
is widely understood to be an inborn feature of supersymmetric
extended objects (for a recent review, see Ref.~\cite{ivanov}).
Exhibiting local kappa invariance, conventional $p$--brane models
enjoy the feature of breaking half of the target space global
supersymmetry. Viewed differently, the PBGS concept can be exploited to
construct superbrane actions in a static gauge~\cite{ik}, the technical tool
here being the method of nonlinear realizations~\cite{coset}.

Recently, there has been growing interest in PBGS options other than
the 1/2 breaking~\cite{t1}--\cite{gght}. This is
essentially due to the discovery of the $d=11$ supergravity solutions
preserving 1/4 or 1/8 of the $d=11$ supersymmetry~\cite{t1} and their
subsequent interpretation in terms of intersecting branes. Since
brane--like worldvolume effective actions which would be capable
of describing those solutions are still unknown, it seems interesting
to study point--like models that mimic the exotic supersymmetry breaking
options inherent in the intersecting branes. Such models could share
some characteristic features of the systems of intersecting branes,
much like the ordinary superparticle bears a similarity
to the Green-Schwarz superstring.

In a series of recent papers~\cite{bl}--\cite{zima}
superparticle models exhibiting 3/4 or 1/4 PBGS have been constructed.
In contrast to the conventional superparticle which, like a single superbrane,
preserves half of the target space supersymmetry, these models reveal some
new interesting peculiarities. Following the argument of
Refs.~\cite{bl,bls,zima}, in order to realize 3/4, 1/4 or some
further fractional PBGS options one has to extend the standard $R^{4|N}$
superspace by new central charge bosonic coordinates. In one of the 1/4 PBGS
massive superparticle models of Ref.~\cite{dik} the target superspace
is $R^{7|8}$. The 1/4 breaking of the original $N=8$ supersymmetry
\footnote{Throughout the paper, $N$ denotes the number of independent
{\it real} supersymmetries from the one-dimensional worldline perspective.}
down to $N=2$ manifests itself in the presence of only one complex
$\kappa$-symmetry in the corresponding worldline action. This is
achieved at cost of the explicit breaking of the target space
Lorentz symmetry down to $SO(3)$ symmetry (in the fermionic sector).

In the present paper we continue the analysis of Ref.~\cite{dik}
and study quantum aspects of this particular $N=8 \rightarrow N=2$ model
as a typical example of massive superparticles with 1/4 PBGS.
We first simplify the Lagrangian of Ref.~\cite{dik} by taking
a real slice in the sector of bosonic variables. This does not
change the structure of global and local symmetries, while still
provides us with an example of 1/4 PBGS in an ordinary four--dimensional
Minkowski space--time (with $R^{4|8}$ as the target superspace and
explicitly broken Lorentz symmetry). Prior to quantization,
Hamiltonian analysis
is accomplished in full detail. A subspace of physical variables is
specified. The supersymmetry algebra is shown to acquire
an extra constant central term which appears differently in the
commutation relations of the broken and unbroken supersymmetry
generators. This is typical of the superbranes in the PBGS approach
and allows one to evade the no-go argument of \cite{witten2} in the line
of the general reasoning of \cite{p}.
%\footnote{See \cite{} for a similar
%consideration in the context of supersymmetric quantum mechanics.}
%and so alowsand realize an
%extension of the ordinary angular momentum algebra compatible with
%Jacobi identities.
We quantize the model in two different ways: in a fixed gauge and using
the Gupta-Bleuler method requiring no gauge-fixing.
Both approaches perfectly match.  We obtain a spectrum of eight
complex on-shell states, four bosons and four fermions, which prove
to be in one--to--one correspondence with the space of differential
zero--, one--, two-- and three--forms on the $x$-manifold.
It is worth mentioning that a similar correspondence is known
to hold in one of the versions of  Witten supersymmetric quantum
mechanics~\cite{witten}. The vacuum structure of the
theory is elucidated and shown to be provided by exact harmonic
one--, and two--forms on the manifold.
%In agreement with the superalgebra
%underlying the theory, the generators of the $SO(3)$ rotations do
%not vanish when acting on the vacuum states, thus being spontaneously
%broken.
Finally, we elaborate on the structure of the representations of
the unbroken  $N=2$ supersymmetry acting in a space of the excited states.
This space is shown to contain two $SO(3)$ scalar and one $SO(3)$ vector
supermultiplets.

%\newpage
%\vskip 0.4cm

\section{Classical Hamiltonian analysis}
%\vskip 0.4cm

According to the original formulation of Ref.~\cite{dik}, a superparticle
realizing the $N=8 \rightarrow N=2$ PBGS mechanism propagates in $R^{7|8}$
superspace. The even part of the supermanifold is parametrized
by seven bosonic coordinates ${x^0,x^i,{\bar x}^i }$, $i=1,2,3$.
The model exhibits an $N=8$ rigid space-time supersymmetry, as well as a local
$\kappa$--invariance with one complex parameter. It is noteworthy that,
without spoiling the symmetry structure, one can reduce the model to
the real subspace $x^i={\bar x}^i$, after which the bosonic coordinates
can be regarded to parametrize the usual four--dimensional flat Minkowski
space. Since this reduction does not invalidate the basic features of the
problem we are dealing with, but considerably simplifies the analysis, in the
rest of the paper we shall concentrate just on this ``real slice'' of the
original model. Its dynamics is governed by the action functional
with a Wess-Zumino term $\sim m$
\bea\label{action}
S=\int d\tau \left\{ \frac {1}{2\,e} \left(-\Pi^0\Pi^0 + \Pi^i\Pi^i\right)
-{1\over 2}\,e\,m^2 +im\left(\theta\dot{\bar{\theta}} -
\psi^i\dot{\bar{\psi}}{}^i \right) \right\}~,
\eea
where
\be
\Pi^0 ={\dot x}^0 +
{\textstyle{\frac i2}} \t \dot{\bt}+{\textstyle{\frac i2}} \bt \dot\t
+{\textstyle{\frac i2}} \p^i \dot{{\bp}}{}^i+
{\textstyle{\frac i2}} {\bp}^i {\dot\p}^i~, \quad
\Pi^i = \dot x^i +i\p^i \dot\t +i{\bar\p}^i \dot{\bar\t}
\ee
and $\t,\p^i$ are four complex fermions parametrizing the odd sector of
the model.

Apart from conventional $\tau$-reparametrizations, the action \q{action}
is invariant under the local $\kappa$--transformations
\bea\label{kappa}
&& \d \t= \kappa,  \quad \delta x^i =-i\p^i \d\t -i{\bar\p}^i \delta {\bar\t}
,\quad
\d \p^i= \frac{\Pi^i \d \bt}{{\Pi}^0 +me},\nonumber\\[2pt]
&& \d x^0 =-{\textstyle{\frac i2}} \t \d \bt-{\textstyle{\frac i2}} \bt \d \t-
{\textstyle{\frac i2}} \p^i \d {\bp}^i-{\textstyle{\frac i2}} {\bp}^i \d \p^i,
\quad \d e = \frac{2ie (\d\t \dot{\bt} +\d\bt \dot\t)}
{{\Pi}^0 +me}.
\eea
Here, $\kappa(\tau)$ is a complex Grassmann parameter.
The action is also invariant under the rigid $x^0, x^i$ translations
extended by the supertranslations with eight real parameters
(or four complex ones $\epsilon^i$, $\epsilon $)
\bea\label{broken}
&& \d \p^i =\e^i, \quad
\d x^0 = -{\textstyle{\frac i2}} \e^i {\bp}^i-{\textstyle{\frac i2}}
{\bar\e}^i \p^i, \quad
\d x^i =-i\e^i \t -i{\bar\e}^i \bar\t~,
\eea
\be\label{unbroken}
\d\t=\e, \quad \d x^0 =-{\textstyle{\frac i2}} \e\bt
-{\textstyle{\frac i2}} \bar\e \t.
\ee
The algebra of the corresponding quantum Noether generators is given below in
Eq.~(\ref{algebra}). Besides, the action \q{action} enjoys a global $SO(3)$
symmetry acting as rotations in the vector index $i$. As distinct from
the standard massive superparticles with a Wess-Zumino term
\cite{1/2,towns,evans} corresponding to 1/2 PBGS, the full $d=4$ Lorentz symmetry
is explicitly broken in \q{action} and is restored only in the limit of
vanishing fermions. One more distinction is that the fermionic variables
are split into a singlet and triplet of the group $SO(3)$, like $x^0, x^i$,
while in the case of $1/2$ superparticles they are in a spinor
representation of the space-time group. In this respect the considered
model resembles a spinning particle where both fermionic and bosonic fields
are space-time vectors. This analogy, however, is rather far-fetched,
since no manifest space-time supersymmetry is present in the spinning
particle.\footnote{An interplay between a $d=4$ spinning particle
and one of the 1/4 PBGS models of Ref. \cite{dik} was studied in \cite{zima}.}
It is also worth noting that the algebra of the global supersymmetry \q{broken},
\q{unbroken} is a truncation of the most general extension of
the standard $N=2, d=4$ ($N=8, d=1$) superalgebra by tensorial
``central charges'' \cite{ferr,gh}, with $P_0, P_i$ being combinations
of the standard $d=4$ translation generators and the central charge
ones \cite{dik}. One more symmetry of \q{action} is the invariance under
phase $U(1)$ transformations of the fermionic variables ($\theta$ and
$\psi^i$ have opposite $U(1)$ charges).

It has to be stressed that, although the manifest Lorentz covariance is
missing in the model under consideration, we can still treat the variable
$x^0$  as a time coordinate in the target space. The corresponding momentum
then specifies the energy
\be
p_0=-p^0=-E,
\ee
with $\eta_{nm}=diag~(-,+,+,+)$. In support of this assertion, the mass
shell condition still holds in the model~\cite{dik}
(see Eq.~(\ref{mshell}) below). The Lorentz invariance
gets broken in the sector of Fermi variables only. Curiously enough,
the situation resembles what happens in the $N=2$ string theory, where
the $U(1)$ current of the $N=2$ superconformal algebra is
constructed out of fermionic fields, which is known to break the full
Lorentz group $SO(2,2)$ down to the subgroup $U(1,1)$ \cite{N2}.
%(see also the discussion in \cite{bg}).

As is well known, the presence of local symmetries is characteristic
of a constrained dynamical system which requires a special care in
quantization. Following Dirac's recipe, in the Hamiltonian framework
one finds five primary constraints
\bea\label{primcon}
&& A \equiv p_\t -{\textstyle{\frac i2}}(p_0 +m) \bt -i p^i \p^i =0,
\quad  \bar A = -p_{\bt} +{\textstyle{\frac i2}}(p_0 +m) \t +i
p^i {\bp}^i =0, \nonumber\\[2pt]
&& A_{i} \equiv  p_{\p i} -{\textstyle{\frac i2}}(p_0 -m) {\bp}_i =0,
\quad  {\bar A}_{i} = -p_{\bp i} +{\textstyle{\frac i2}}(p_0 -m) \p_i =0,
\quad p_e =0,
\eea
while the complete canonical Hamiltonian reads
\be \label{canham}
H=p_e \l_e +A \l_\t -\bar A \l_{\bt} + A^i \l_\p^i -
{\bar A}^i \l_{\bp}^i +
{\textstyle{\frac 12}} e(m^2 -p_0 p_0 + p^i p^i ).
\ee
Here $(p_\t,p_0, p_i,p_{\p^i}, p_e)$ stand for the momenta canonically
conjugate to the variables $(\t,x^0,$
$x^i,\p^i,e)$ and $\l_{e}$, etc, are the Lagrange multipliers. The canonical
brackets read
\be \label{canon}
\{x^0, p_0 \} =\{e, p_e \} =1~, \; \{x^i, p^k\} = \delta^{ik}~, \;
\{p_\theta, \theta\} = \{p_{\bar\theta}, \bar\theta\} = 1~, \;
\{p_\psi^i, \psi^k\} = \{p_{\bar\psi}^i, \bar\psi^k\}
= \delta^{ik}~.
\ee
Given the Hamiltonian \q{canham}, conservation of the primary constraints
implies one secondary constraint
\be\label{mshell}
M \equiv m^2 -p_0 p_0+p^i p^i   =0,
\ee
and specifies some of the Lagrange multipliers
\bea\label{lm}
&& (p_0 -m) \l_{\p i} +p_i \l_{\bt}=0, \quad
(p_0 -m) \l_{\bp i} +p_i \l_\t=0, \nonumber\\[2pt]
&&  (p_0 +m) \l_\t  +p^i \l_{\bp}^i=0, \quad
(p_0 +m) \l_{\bt}  +p^i \l_{\p}^i=0.
\eea
Hereafter, we eliminate the canonically conjugate pair of
non-dynamical variables $p_e$ and $e$ from the consideration
in a standard way (by fixing the gauge $e = const$, after which
the constraint $p_e$ becomes second-class and $p_e$ can be
removed altogether by passing to the appropriate Dirac bracket).

Aiming at the construction of a quantum mechanical description of the
system at hand, in the following we shall restrict ourselves to the upper
shell of the massive hyperboloid~(\ref{mshell})
\be\label{uppershell}
p^0 = E \ge m~, \;\;\mbox{or} \;\; p_0 \le -m~, \;\; \Rightarrow \;\;
p_0 - m \neq 0~,
\ee
thus omitting configurations with negative energy. Under this assumption,
Eqs.~(\ref{lm}) determine the value of $\l_\p$ and $\l_{\bp}$, while still
leave $\l_\t, \l_{\bt}$ arbitrary. The latter fact signals the presence of
two first-class fermionic constraints in the formalism.
The separation of the constraints into the first- and second-class ones
becomes more transparent after the simple redefinition
\be\label{firstclass}
A \rightarrow A'=A + \frac {1}{(p_0 -m)} p^i {\bar A}^i.
\ee
In the new basis the full set of the canonical Poisson brackets between
the basic constraints is as follows
\bea
&& \{A', \bar A' \} = -i\frac{1}{(p_0-m)} (m^2- p_0p_0 + p^ip^i ) \approx 0 ~, \nn
&& \{A', A^i \} = \{A', \bar A^i\} = \{\bar A{}', A^i \} = \{\bar A{}',
\bar{A}^i \} = 0~,  \nonumber \\
&& \{ A^i, \bar{A}^k \} = i(p_0 - m)\delta^{ik}~, \quad \{A^i, A^k\}
= \{\bar{A}^i, \bar{A}^k \} = 0~. \label{canon2}
\eea
We see that $A', \bar A', M$ are first-class, while
$A_i,{\bar A}_i$ are second-class
\bea
&&\underline{\mbox{First class:}} \quad A' \approx 0~, \;\bar A' \approx 0~, \;
M \approx 0~, \label{first} \\
&&\underline{\mbox{Second class:}} \quad A^i \approx 0~, \; \bar A^i \approx 0~.
\label{second}
\eea
With respect to the corresponding Dirac bracket the constraints
$A', \bar A', M$ generate, respectively, the complex $\kappa$-symmetry and
$\tau$-reparametrizations. Such a bracket is easy to construct, but we postpone
giving its explicit form  until fixing a gauge with respect to
the $\kappa$-symmetry.
%The constraints .
%among the latter reads
%\bea\label{t}
%&& \{ A',\bar A' \}=-{\textstyle{\frac {i}{(p_0 -m)}}}
%(m^2  +p^i p^i -p_0 p_0) \approx 0.
%\eea

In the next Sections we shall quantize the theory in two different ways,
either eventually leading to the same spectrum of physical states. One of them
is the Gupta-Bleuler quantization which can be
performed with all local symmetries being kept manifest. Another one
involves removing, prior to quantization, some irrelevant
unphysical variables by fixing proper gauges with respect to the local
symmetries. In this case one should necessarily deal with Dirac
brackets. \footnote{A quantization of massless superparticle with
the gauge-fixed $\kappa$-symmetry was accomplished in \cite{evans2}.}
In the rest of this Section we describe the Hamiltonian formalism
along the lines of the second approach.

We impose the gauge conditions
\be
\t=0, \quad \bt=0~. \label{gauge}
\ee
Conservation of the gauge fully specifies the value of the
remaining independent Lagrange multipliers in \q{canham}, \q{lm}
\be
\l_\t = \l_{\bt} =0~.
\ee
As usual, the gauge-fixing
conditions, together with the former first-class constraints
$A', \bar{A}' $, should now be treated as second-class
constraints extending the set \q{second}. The Dirac bracket then has to be
used for the remaining variables. For the case at hand it is defined by
\bea
&& {\{ B,C \}}_D =\{ B,C \} +\frac{i}{(p_0 -m)}\{B,\t \} M \{\bt, C \} +\frac{i}{(p_0-m)}\{B,\bt \} M \{\t, C \}
\nonumber\\[2pt]
&& \qquad  \qquad \quad -
\{B,\t \} \{A', C \}+\{B,\bt \} \{{\bar A}', C \}-\{B,A' \} \{ \t, C \}+
\{B,{\bar A}' \} \{ \bt, C \}
\nonumber\\[2pt]
&& \qquad \qquad \quad + \frac {i}{(p_0 -m)}
\{B,A_i \} \{{\bar A}_i, C \} +
\frac {i}{(p_0 -m)}  \{B,{\bar A}_i \} \{ A_i, C \}, \label{dirac}
\eea
where $M$ is defined in \q{mshell}. Being evaluated in the coordinate
sectors, \q{dirac} gives
\bea\label{fundbr}
&&
{\{x^0, \p^i \}}_D =- \frac {1}{2(p_0 -m)} \p^i~,
\quad {\{x^0, {\bp}^i \}}_D =-\frac {1}{2(p_0 -m)} {\bp}^i~,
\nonumber\\[2pt]
&& {\{ x^0, p_0 \}}_D =1, \quad {\{ x^i, p^j \}}_D =\d^{ij}~,
\quad {\{ \p^i, {\bp}^j \}}_D=- \frac {i}{(p_0 -m)} \d^{ij}~,
\eea
with all other brackets vanishing.

Let us dwell on the issue of global
supersymmetry in the reduced phase space which might give us a filling
of what type of symmetries one has to expect at the quantum level.
In the chosen gauge, the equations of motion take their free form
\be
{\dot x}^0 = -p_0~, \quad {\dot x}^i =p^i~, \quad {\dot p}_0 =0~,
\quad {\dot p}^i =0~, \quad {\dot\p}^i =0~.
\ee
Then, recalling the original transformation laws \q{kappa} - \q{unbroken},
one finds that six of the global supersymmetries are now realized as
\bea\label{broken1}
\d \p^i =\e^i~, \quad
\d x^0 = -{\textstyle{\frac i2}} \e^i {\bp}^i-{\textstyle{\frac i2}}
{\bar\e}^i \p^i~, \quad \d x^i=0~.
\eea
Two remaining supersymmetries \q{unbroken} are now modified by
a compensating $\kappa$--transformation \q{kappa} chosen so as to
preserve the gauge \q{gauge}
\bea\label{unbroken1}
 \d \p^i= \frac {1}{(p_0 -m)} \bar\e p^i~,\quad
\d x^i =i\p^i \e +i {\bar\p}^i \bar\e~,
\quad  \d x^0 =- \frac {i}{2(p_0 -m)} p^i(\p^i \e +
{\bar\p}^i \bar\e)~.
\eea
As a typical feature of the canonical formalism, the action of
some symmetry generator $Q^i$ is defined via the Dirac bracket as follows
\be
\d B = i\{ B, Q^i \}_D \e^i +
i\{ B, {\bar Q}^i \}_D {\bar\e}^i~. \label{transfD}
\ee

Aiming at quantization of the system as the eventual goal, we first diagonalize
the brackets~(\ref{fundbr}) by redefining the fermionic fields
\bea\label{newfields}
&& \p^i \rightarrow \p^{'i} =\p^i \sqrt{p_0 -m}~, \quad
{\bp}^i \rightarrow {\bp}^{'i} ={\bp}^i \sqrt{p_0 -m}~, \nonumber\\[2pt]
&& \{ x^0, p_0 \}_D =1~, \quad \{ x^i, p_j \}_D ={\d^i}_j~, \quad
 \{ \p^{'i}, {\bp}^{'j} \}_D =-i \d^{ij}~,
\eea
which makes the passage to a quantum description straightforward.
In the new basis the supersymmetry generators take the form
(hereafter, we omit primes on the new fields)
\be
Q^i ={\bp}^i \sqrt{p_0-m}~, \quad {\bar Q}^i =\p^i \sqrt{p_0-m}~,
\quad Q= \frac {1}{\sqrt{p_0 -m}} \p^i p^i~,
\quad \bar Q = \frac {1}{\sqrt{p_0 -m}} {\bp}^i p^i~. \label{susygen}
\ee
One should add to these generators also the generators of $SO(3)$ rotations
\be\label{angular}
J_i=\e_{ijk} x^j p^k -i\e_{ijk} {\bp}^j \p^k~.
\ee
We can now write the full closed superalgebra at once in the quantum case,
making the standard replacement of the Dirac bracket by the graded
commutator
\be
\{ \quad \}_D \Rightarrow -i \left( \{ \quad \}~, \;
[ \quad] \right)~, \label{repl}
\ee
where the anticommutator is chosen for the bracket between two fermionic
operators.

The full quantum algebra including \q{susygen} together with the translation
generators $p_0 = -i\partial/\partial x^0 , p_i = -i\partial/\partial x^i $
and the generators of $SO(3)$ rotations \q{angular} then reads
\bea\label{algebra}
&& \{ Q^i, {\bar Q}^j \} =(p_0 -m) \d^{ij}~, \quad \{Q, \bar Q \}=
(p_0 +m)+ \frac {1}{(p_0 -m)} (m^2 -p_0 p_0+ p^i p^i )~,
\nonumber\\[2pt]
&& \{Q, Q^i \} =p^i~, \quad \{\bar Q, {\bar Q}^i \} =p^i~, \nonumber\\[2pt]
&& [J_i, J_j] =i\e_{ijk} J^k~, \; [J_i, p_j] =i\e_{ijk} p^k~,
\; [J_i, Q_j] =i\e_{ijk} Q^k,~ \; [J_i, {\bar Q}_j] =i\e_{ijk} {\bar Q}^k~.
\eea
Other (anti)commutators prove to vanish. One can directly check that these
generators, by the general rule \q{transfD} (with the replacement \q{repl}),
yield for the target superspace coordinates just the supersymmetry
transformations \q{broken1}, \q{unbroken1}, translations and standard
$SO(3)$ rotations.
%It is straightforward to check
%that Jacobi identities hold for the superalgebra provided the
%well known cyclic identity
%\be
%\e_{ijk} {\e^k}_{lm} +cycle(ijl)=0.
%\ee

Worth noting is the appearance of the constant central charge $\pm m$ in the
anticommutators $\{Q, \bar Q \}$ and $\{Q_i, \bar Q_j \}$
and the weakly vanishing term in the first anticommutator.
The latter property is typical for gauge-fixed theories. Recall that
the equation
\be
M = m^2-p_0 p_0 +p^i p^i  =0 \label{Mshell}
\ee
is the only first-class constraint remaining in the formalism.
Following Dirac's method one should require it to vanish on physical states.
When restricted to the physical subspace, the algebra~(\ref{algebra})
thus acquires its rigorous form. From the structure of the
algebra one can also infer that in the realization on the states
the generators $Q^i, {\bar Q}^i$ should correspond to the spontaneously
broken symmetries (recall that by assumption $p_0-m \ne 0$),
while $Q, \bar Q$ can be chosen to be unbroken and so annihilating
the vacuum. The appearance of the constant central charge $m$ with
opposite signs in the anticommutators of broken and unbroken supersymmetries
ensures evading the arguments of \cite{witten2} against the possibility of
partial breaking, in accord with the generic reasoning of ref. \cite{p}
(it is applicable to any superbrane theory).

Finally, it is important to stress that it is the mass--shell
condition \q{Mshell} that allowed one to construct an
$N=8$ supersymmetry algebra out of the operators at hand, with six
supersymmetries being broken. Similar to other superbrane-like models,
the partial breaking thus holds at the free theory level, without need
to introduce a potential of a specific shape, as it takes place
in the standard non-relativistic supersymmetric quantum
mechanics~\cite{witten2,fub,ikp}.

\setcounter{equation}{0}
\section{Quantization in a fixed gauge and the vacuum \break
structure}
%\vskip 0.4cm
Let us proceed to the more detailed exposition of the quantization procedure.
After replacing the Dirac brackets by (anti)comutators according to the rule
\q{repl} we represent the fermionic coordinates by means of conventional
creation--annihilation operators
\bea\label{ladders}
\p^i \rightarrow a^i, \quad  {\bar \p}^i \rightarrow a^{i +}, \quad
\{ a^i, a^{j +}  \}= \delta^{ij}.
\eea
For the bosonic operators we keep the ordinary coordinate
representation, with
\be
p_0 = -i\frac{\partial}{\partial x^0},  \quad
p_i = -i \frac{\partial}{\partial x^i}~. \label{ppquant}
\ee

Given a single pair of fermionic  operators, a convenient
matrix representation is \cite{fub,affl}
$$
\begin{array}{lll}
a=
\left(\begin{array}{cccc}
0 & 1 \\
0 & 0\\

\end{array}\right),

\qquad
a^{+}=
\left(\begin{array}{cccc}
0 & 0 \\
1 & 0\\
\end{array}\right)~,
\end{array} \quad \{a, a^+ \} = 1~.
\eqno{(3.2)}
$$
\addtocounter{equation}{1}
A representation space is trivially constructed and consists of a vacuum
state and a single filled state
$$
\begin{array}{lll}
\arrowvert 0 \rangle=
\left(\begin{array}{cccc}
1 \\
0\\
\end{array}\right),
\qquad
\arrowvert \uparrow \rangle= a^+ \arrowvert 0 \rangle=
\left(\begin{array}{cccc}
0 \\ 1 \\
\end{array}\right)~.
\end{array}
\eqno{(3.3)}
$$
\addtocounter{equation}{1}
To construct a representation for the triplet~(\ref{ladders}),
it suffices to find a matrix which anticommutes with both $a$ and
$a^{+}$. Such a matrix is readily constructed
$$
\begin{array}{lll}
\tau=[a,a^{+}] =
\left(\begin{array}{cccc}
1 & 0 \\
0 & -1\\
\end{array}\right),
\qquad {\tau}^{2}=1,
\end{array}
\eqno{(3.5)}
$$
\addtocounter{equation}{1}
and the sought representation is then given by $8\times 8$ matrices
\bea
&& a_1=a\times 1_2 \times 1_2, \qquad a_2=\tau \times a \times 1_2,
\qquad a_3=\tau \times \tau \times a, \nonumber\\[2pt]
&&
a_1^{+}=a^{+}\times 1_2 \times 1_2, \qquad a_2^{+}=\tau \times a^{+}
\times 1_2, \qquad a_3^{+}=\tau \times \tau \times a^{+}. \label{8repr}
\eea
It is noteworthy that the properties of the $\tau$ operator allow one
to identify it with a $Z_2$--grading operator (sometimes referred to as the
Klein operator) acting in the Hilbert space (see, for e.g.,~\cite{gen,ply}).
In particular, the eigenstates corresponding to the  eigenvalue $+1$ of
this operator are identified with bosonic states (for the simplest case
of one pair of the creation-annihilation operators there is only one such
state,  $\arrowvert 0 \rangle$), while those corresponding to the
eigenvalue $-1$ are identified with fermions (in the simplest case
the only fermionic state is $\arrowvert \uparrow \rangle$).

In accord with the realization \q{8repr}, the representation space
of the full algebra is eight-dimensional~\footnote{The direct products of
the states
$\arrowvert 0 \rangle$ and  $\arrowvert \uparrow \rangle$
amount to usual eight component
columns \\
$\arrowvert 0 \rangle \times
\arrowvert 0 \rangle \times \arrowvert 0 \rangle=
$$
\begin{array}{lll}
\left(\begin{array}{cccc}
1  \\
0 \\
0\\
\vdots
\end{array}\right),
\quad
\arrowvert \uparrow \rangle \times
\arrowvert 0 \rangle \times \arrowvert 0 \rangle=
\left(\begin{array}{cccc}
0  \\
1 \\
0\\
\vdots
\end{array}\right),
\quad \dots \quad,
\arrowvert \uparrow \rangle \times
\arrowvert \uparrow \rangle \times \arrowvert \uparrow \rangle=
\left(\begin{array}{cccc}
\vdots  \\
0 \\
0\\
1\\
\end{array}\right).
\end{array}
$$
$}
\bea\label{states}
&&
 \qquad \qquad \qquad
\arrowvert 0 \rangle \times \arrowvert 0 \rangle \times
\arrowvert 0 \rangle \otimes \Phi(x),
\nonumber\\[2pt]
&&
\arrowvert \uparrow \rangle \times \arrowvert 0 \rangle \times
\arrowvert 0 \rangle  \otimes \Psi_1 (x), \quad
\arrowvert 0 \rangle \times \arrowvert \uparrow \rangle \times
\arrowvert 0 \rangle   \otimes \Psi_2 (x), \quad
\arrowvert 0 \rangle \times \arrowvert 0 \rangle
\times \arrowvert \uparrow \rangle  \otimes \Psi_3 (x),
\nonumber\\[2pt]
&& \arrowvert 0 \rangle \times \arrowvert \uparrow \rangle \times
\arrowvert \uparrow \rangle  \otimes \Phi_1 (x), \quad
\arrowvert \uparrow \rangle \times \arrowvert 0 \rangle \times
\arrowvert \uparrow \rangle \otimes \Phi_2 (x), \quad
\arrowvert \uparrow \rangle \times
\arrowvert \uparrow \rangle \times \arrowvert 0 \rangle
\otimes \Phi_3 (x), \nonumber\\[2pt]
&& \qquad \qquad \qquad
\arrowvert \uparrow \rangle \times \arrowvert \uparrow \rangle \times
\arrowvert \uparrow \rangle \otimes \Psi (x),
\eea
with $x=(x^0,x^i)$,
and it is the direct sum of two complex $SU(2)$ singlets $\Phi$, $\Psi$ and
two complex $SU(2)$ triplets $\Phi_i$, $\Psi_i$ (total of $8 + 8$
states). It should be stressed that we do not assign the Fermi statistics to any of the $x$--dependent
functions appearing above. The statistics of the states is defined
entirely with respect to the $Z_2$--grading operator
$\tau \times \tau \times \tau$. Thus, we have a single boson in the first
line of Eq.~(\ref{states}), a triplet of fermions in the second line,
a triplet of bosons in the third line and a single fermion in the last line.
Note that this decomposition into fermionic and bosonic states is to some extent
conventional. As the $Z_2$--grading operator one could equally take
$-\tau$ , with respect to which the bosonic states become fermionic
and vice versa. Similarly, the vacuum and filled states in (3.3),
as well as the creation and annihilation operators, alternate their status.
Without loss of generality, in what follows we shall stick to
the first grading.

Of frequent use in the literature are also alternative
representations which deal with either superfields or a more
abstract Fock space (see, e.g,~\cite{fub,ikp}).
%According to such
%a point of view, a general quantum state (superfield) can be chosen to be
%either bosonic (the ground state is fully empty) or fermionic
%(the ground state is completely filled with fermions). As a matter of fact,
%one has to adopt in parallel also a dual description, the mutually dual
%spaces being connected by the supersymmetry generators. In the matrix
%representation we stick to in this section the supersymmetry charges
%do not take a state out of the chosen Hilbert space. Hence, no need
%in a dual space arises. As an alternative,
In the next Section we shall present the superfield Gupta-Bleuler
quantization of the same system and show that it yields
an equivalent spectrum of states.

At first glance, it seems somewhat surprising that a pithy part of the
states~(\ref{states}) is described by purely bosonic functions. Observe,
however, that the four levels in \q{states} are in one--to--one
correspondence with the space of differential zero--, one--, two--
and three--forms on a manifold (the components of a 2--form are defined as
$\Phi_{ij} =\epsilon_{ijk} \Phi^k$ while those of a 3--form as
$\Psi_{ijk}=\epsilon_{ijk} \Psi$). At this stage it seems relevant to
mention that the {\it de Rham} complex of a (curved) manifold,
the space of all  $p$--forms,  can be described within the
framework of supersymmetric quantum  mechanics~\cite{witten}.
This correspondence between fermionic states and $p$-forms is also
reminiscent of K\"ahler's geometric reformulation of spinors and
Dirac equation in terms of differential forms (for a comprehensive review
and further references see Ref.~\cite{becher}).

Armed with these remarks, we now proceed to analyse the vacuum structure of the theory. Most
elegantly this can be done again in terms of differential forms and our
discussion here parallels that of Ref.~\cite{witten}.
Due to the algebra~(\ref{algebra}), the vacuum state of the unbroken
supersymmetry defined by the conditions
\be\label{vacstate}
Q  \arrowvert vac \rangle = \bar Q  \arrowvert vac \rangle=  0,
\ee
necessarily has minimal energy
\be \label{vacen}
p_0+m=0~, \quad \Rightarrow \quad M = p^ip^i -p^0p^0 + m^2 = p^ip^i \equiv
-\Delta~.
\ee
Then the supersymmetry charges $Q$ and $\bar Q$ can be given a natural
geometric interpretation. When acting on a vacuum state they coincide
with the exterior differentiation $d$ and the adjoint exterior
differentiation $\delta$, respectively
\be\label{vaceq}
\bar Q \sim d, \quad Q \sim \delta, \quad d\delta +\delta d = {1\over 2m}\Delta.
\ee
An immediate consequence of \q{vacstate} - \q{vaceq} is that the vacuum state
of the unbroken supersymmetry necessarily involves a harmonic form. Since
$d$ increases the order of a form by one unit while $\delta$ decreases it
by one unit, it suffices to apply the operators $Q$ and $\bar Q$
directly to each level in Eq.~(\ref{states}) (to be more precise, one
has to consider a linear combination of states at a given level)
without need to consider any linear combination of states belonging to
different levels.

On a manifold of trivial topology,
which we assume in this work, one finds the following
solution to Eq.~(\ref{vacstate}) in terms of 0--, or 3--forms
(the first and the fourth levels in Eq.~(\ref{states})):
\be
{\arrowvert vac \rangle}_{(B)}^{(0)} =
\arrowvert 0 \rangle \times \arrowvert 0 \rangle \times
\arrowvert 0 \rangle \otimes e^{-imx^0} \alpha~, \quad
{\arrowvert vac \rangle}_{(F)}^{(0)} =
\arrowvert \uparrow \rangle \times \arrowvert \uparrow \rangle \times
\arrowvert \uparrow \rangle \otimes e^{-imx^0}\beta~, \label{triv}
\ee
where $\alpha, \beta$ are some constants. These vacua are not
too interesting. Indeed, on them
\be
p^i{\arrowvert vac \rangle}_{(B)}^{(0)} =
p^i{\arrowvert vac \rangle}_{(F)}^{(0)} = 0~,
\ee
and, as follows from the (anti)commutation relations \q{algebra}, the
$(Q, \bar Q, p_0 + m)$ and $(Q^i, \bar Q^i, p_0 -m)$ supersymmetries decouple
from each other. First supersymmetry is unbroken, while the second one
is totally broken. The only Goldstone excitations are expected to be
the complex Volkov-Akulov \cite{VA} Goldstone fermions associated
with the generators $Q^i$, or $\bar Q^i$. The action of the latter
on \q{triv} produces a ring of ground states, every state possessing
the minimal energy \q{vacen} and being a singlet of the $Q, \bar Q$
supersymmetry. The holomorphic set $Q^i$ annihilates
${\arrowvert vac \rangle}_{(F)}^{(0)}$, while the conjugated set
vanishes on ${\arrowvert vac \rangle}_{(B)}^{(0)}$.

Thus these vacua and the related sector of the full space of states
do not correspond to the symmetry structure of the 1/4 PBGS superparticle
of Ref. \cite{dik}. Indeed, in the latter case the translations $p^i$
should also be necessarily broken, with the associated Goldstone excitations
as the transverse superparticle coordinates.

The vacua with the desirable properties arise as solutions of
Eqs.~(\ref{vacstate}) for the second and third levels in \q{states}.
For the 1--forms (the second level in \q{states}) Eqs.~(\ref{vacstate})
amount to
\be
p^i\Psi^i=0, \qquad \partial^{[i}\Psi^{j]}=0 \rightarrow
\Psi^i= e^{-i mx^0}\,p^i~\Sigma (\vec{x}), \quad \Delta \Sigma=0,
\ee
and the general structure of the corresponding fermionic vacuum state is
\be
{\arrowvert vac \rangle}_{(F)}=a^{i +}
\arrowvert 0 \rangle \times \arrowvert 0 \rangle \times
\arrowvert 0 \rangle \otimes p^i~ e^{-imx^0}~ \Sigma(\vec{x}), \quad
\Delta \Sigma =0~. \label{fermVac}
\ee
For the 2--forms (the third level in~(\ref{states})) Eqs.~(\ref{vacstate})
can be analysed in the same spirit, yielding a bosonic vacuum state
\be
{\arrowvert vac \rangle}_{(B)}=a^{i +} a^{j +} \epsilon_{ijk}
\arrowvert 0 \rangle \times \arrowvert 0 \rangle \times
\arrowvert 0 \rangle \otimes p^k~ e^{-imx^0}~ \Omega(\vec{x}), \quad
\Delta \Omega =0~. \label{bosVac}
\ee
We are led to neglect in $\Sigma(\vec{x})$, $\Omega(\vec{x})$ zero modes
$\sim x^i$, since the corresponding pieces belong to the ring of
``trivial'' vacua \q{triv}.

It is straightforward to check that none of the generators $Q^i$ and
${\bar Q}^i$ annihilate the vacuum states defined in this way;
these generators rather produce one or another multiplet of
the unbroken $N=2$ supersymmetry.
The resulting  states certainly do not belong to the ring
of vacua (i.e. do not obey eqs. \q{vacstate}) in view of the
anticommutation relations \q{algebra} and the important property
\be
p^i{\arrowvert vac \rangle}_{(F,B)} \neq 0.
\ee
In full agreement with the classical consideration \cite{dik}, one
concludes that these six supersymmetries are spontaneously broken together
with three transverse translations, i.e., this vacuum structure and the
associated sector of the space of quantum states precisely match
the ``real slice'' of the 1/4 PBGS superparticle of \cite{dik}
with which we started in Section 2.

It remains to discuss the generators of the
$SO(3)$ rotations. Making use of the explicit
representation~(\ref{angular}) one can readily verify the relations
\bea
&& J_i {\arrowvert vac \rangle}_{(F)}=a^{j +}
\arrowvert 0 \rangle \times \arrowvert 0 \rangle \times
\arrowvert 0 \rangle \otimes e^{-imx^0}p^j \Sigma_i (\vec{x}), \quad
\Sigma_i(\vec{x}) = \e_{ijk} x^j p^k \Sigma (\vec{x})~,
\nonumber\\[2pt]
&& J_i {\arrowvert vac \rangle}_{(B)}=a^{j +} a^{k +} \e_{jkl}
\arrowvert 0 \rangle \times \arrowvert 0 \rangle \times
\arrowvert 0 \rangle \otimes e^{-imx^0} p^l \Omega_i (\vec{x}), \quad
\Omega_i(\vec{x}) =\e_{ijk} x^j p^k \Omega(\vec{x}). \nn
\eea
Since the operators $J_i$ do not annihilate these vacuum states, but rather
produce new vacua of the same sort, generically they are
spontaneously broken. Note that they are vanishing on the ``trivial''
vacua \q{triv}, indicating that $SO(3)$ is unbroken in the sector
corresponding to two decoupled supersymmetries. However, it can be
chosen unbroken in the considered ``1/4 PBGS superparticle'' sector as well,
provided one selects some subclass in the set of vacua \q{fermVac},
\q{bosVac}. Indeed, the relations
\be
\Sigma_i = \Omega_i = 0
\ee
hold on the spherically-symmetric solutions of the Laplace equation:
\be
\Sigma(\vec{x}) = const_1 + {1\over \vert \vec{x}\vert}~, \quad \Omega(\vec{x}) =
const_2 + {1\over \vert \vec{x}\vert}
\ee
(actually, $const_1$ and $const_2$ drop out from the
corresponding subset of the vacua \q{fermVac} and \q{bosVac}).

In the end of the next Section we shall briefly discuss how this
vacuum PBGS structure is related to the standard treatment of the partial
breaking of supersymmetry in the field theory models, and in which precise
sense it implies the presence of the appropriate Goldstone excitations in
the spectrum.

Finally, let us comment on the structure of the representation of the
$N=2$ unbroken supersymmetry which acts in a space of the ``excited''
(i.e., with $E = -p_0 > m$) states.
Since in this case
\be
{\vec{p}}{\;}^2 = p^i p^i \ne 0,
\ee
for the fermionic states from the second line in Eq.~(\ref{states}) one can
use the decomposition
\be
\Psi^i= \left(\d^{ij} - \frac{p^i p^j}{{\vec{p}}{\;}^2}\right) \Psi^j
+ \frac{p^i p^j}{{\vec{p}}{\;}^2} \Psi^j \equiv
\Psi_{\bot}^i +p^i~ \Upsilon, \qquad p^i\Psi_{\bot}^i=0.
\ee
Analogously, the bosonic states from the third line in Eq.~(\ref{states})
can be represented as
\be
\Phi^i=\left(\d^{ij} - \frac{p^i p^j}{{\vec{p}}{\;}^2}\right) \Phi^j
+ \frac{p^i p^j}{{\vec{p}}{\;}^2} \Phi^j \equiv
\Phi_{\bot}^i +p^i~ \Xi, \qquad p^i\Phi_{\bot}^i=0.
\ee
With a simple inspection one can further verify that
at each level the states
\be
\arrowvert 0 \rangle \times \arrowvert 0 \rangle \times
\arrowvert 0 \rangle \otimes \Phi(x), \quad
a^{i +} \arrowvert 0 \rangle \times \arrowvert 0 \rangle \times
\arrowvert 0 \rangle \otimes p^i~ \Upsilon(x), \label{a}
\ee
\be
a^{i +} a^{j +} \epsilon_{ijk}
\arrowvert 0 \rangle \times \arrowvert 0 \rangle \times
\arrowvert 0 \rangle \otimes \Phi_{\bot}^k (x),
\quad
a^{i +}
\arrowvert 0 \rangle \times \arrowvert 0 \rangle \times
\arrowvert 0 \rangle \otimes \Psi_{\bot}^i (x), \label{b}
\ee
\be
a^{i +} a^{j +} \epsilon_{ijk}
\arrowvert 0 \rangle \times \arrowvert 0 \rangle \times
\arrowvert 0 \rangle \otimes p^k~ \Xi (x),
\quad
a^{i +} a^{j +} a^{k +} \epsilon_{ijk}
\arrowvert 0 \rangle \times \arrowvert 0 \rangle \times
\arrowvert 0 \rangle \otimes \Psi (x), \label{c}
\ee
form irreducible multiplets of the unbroken $N=2$ supersymmetry.
One thus concludes that the space of the excited states is a direct sum
of these three on-shell representations of one-dimensional
$N=2$ supersymmetry, involving, respectively, $(2 +2)$, $(4 + 4)$ and
$(2+2)$ independent real components. The rest of $N=8$ supersymmetry generators,
$Q^i, \bar Q^i, $ mix these $N=2$ multiplets with each other,
combining them into an irreducible on-shell multiplet of the
full supersymmetry.
%\vspace{0.4cm}

\setcounter{equation}{0}
\section{Gupta-Bleuler quantization}
%\vspace{0.4cm}

%We divide all constraints into two conjugated sets
%\be\label{sets}
%A_I=\left\{ {\bar A}, A^i \right\} \; , \; A_{II}=\left\{ A, {\bar A}^i \right\} \;.
In the GB quantization (see, e.g., \cite{luk}) one represents the wave function
by a complex superfield $\varphi$,
\be
\varphi=\varphi( x^0,x^i,\theta,\bar\theta, \psi^i, \bar\psi{}^i ) \; ,
\ee
and imposes on it all the first-class constraints \q{first} and half of the
second-class constraints \q{second} (without passing to Dirac bracket).
%$\left\{ A', {\bar A}{}'\right\}$ \q{firstclass} and half of
%the second class constraints.

We shall enforce these constraints in two steps:
\begin{enumerate}
\item Off-shell constraints: $A^i \varphi=0, \; {\bar A{}'} \varphi=0$~;
\item On-shell  constraints: $A' \varphi=0, \; (m^2 -p_0^2 +p^i p^i)\varphi=0$~.
\end{enumerate}

We replace the momenta by differential operators
\be
p_\theta\rightarrow i\frac{\partial}{\partial\theta},\;
{\bar p}_\theta\rightarrow i\frac{\partial}{\partial\bar\theta},\;
p^i_\psi\rightarrow i\frac{\partial}{\partial\psi^i},\;
{\bar p}{}^i_\psi\rightarrow i\frac{\partial}{\partial\bar\psi{}^i}.
\ee
after which the off-shell constraints take the form
\be
D_i \varphi = 0~, \quad \left(\bar D + p^i\bar{\psi}^i \right)\varphi = 0~,
\label{off}
\ee
where
\bea
&& D_i = \frac{\partial}{\partial \psi^i} - {1\over 2}(p_0 - m)\bar \psi_i~,
\quad \bar D_i = - \frac{\partial}{\partial \bar\psi^i} +
{1\over 2}(p_0 - m)\psi_i~, \nn
&& D = \frac{\partial}{\partial \theta} - {1\over 2}(p_0 + m)\bar \theta~,
\quad
\bar D = - \frac{\partial}{\partial \bar\theta} +
{1\over 2}(p_0 + m)\theta~. \label{defDsp}
\eea
The solution of \q{off} reads
\bea\label{off1}
\varphi&=& u+{\bar\psi}{}^i{\bar\rho}{}^i+{\bar\psi}{}^{2i}{\bar v}{}^i +{\bar\psi}{}^3{\bar\eta} \nn
    && + \frac{1}{2}(p_0-m)\psi^i\left( {\bar\psi}{}^i u - \epsilon^{ijk}{\bar\psi}{}^{2j}{\bar\rho}{}^k +
        {\bar\psi}{}^3{\bar v}{}^i \right) \nn
    && -\frac{1}{4}(p_0-m)^2 \psi^{2i} \left( {\bar\psi}{}^{2i} u +{\bar\psi}{}^3\right)-
       \frac{1}{8}(p_0-m)^3\psi^3{\bar\psi}{}^3 u \; .
\eea
Here
\be
\psi^{2i}\equiv \frac{1}{2}\epsilon^{ijk}\psi^j\psi^k\;,\;
\psi^{3}\equiv \frac{1}{6}\epsilon^{ijk}\psi^i\psi^j\psi^k\;,
\ee
and the $\bar\psi$-monomials are defined by the same formulas. The
superfields $\left\{ u,{\bar\rho}{}^i,{\bar v}{}^i,{\bar\eta}\right\}$
depend only on $\left\{   x^0,x^i,\theta,\bar\theta\right\}$ and obey
the constraints
\be
{\bar D}u=0\;,\;{\bar D}{\bar\rho}{}^i=-p^i u\; ,\;
{\bar D}{\bar v}{}^i =\epsilon^{ijk}p^j{\bar\rho}{}^k\;,\;
  {\bar D}{\bar\eta}=-p^i{\bar v}{}^i \;.
\ee
In terms of the components fields, the solution of the off-shell
constraints reads
\bea\label{components}
&& u=u_0+\theta\bar\xi -\frac{1}{2}\theta\bar\theta (p_0+m)u_0 \; , \nn
&& {\bar\rho}{}^i= {\bar\rho}{}^i_0+\theta{\bar\phi}{}^i-\bar\theta (p^i u_0)+
   \theta\bar\theta \left(p^i{\bar\xi}-\frac{1}{2}(p_0+m){\bar\rho}{}^i_0\right) \;, \nn
&& {\bar v}{}^i={\bar v}{}^i_0+ \theta{\bar\zeta}{}^i +\bar\theta \epsilon^{ijk}p^j{\bar\rho}{}^k_0+
   \theta\bar\theta\left( -\epsilon^{ijk}p^j{\bar\phi}{}^k-\frac{1}{2}(p_0+m){\bar v}{}^i_0\right)\;, \nn
&& \bar\eta=\bar\eta_0+\theta\bar\omega -\bar\theta (p^i {\bar v}{}^i_0)+
  \theta\bar\theta\left( p^i{\bar\zeta}{}^i -\frac{1}{2}(p_0+m)\bar\eta_0\right) \;.
\eea
Thus, off shell we have 8 complex bosonic fields
\be
u_0\;,\;{\bar\phi}{}^i\;,\;{\bar v}{}^i_0\;,\; {\bar\omega}
\ee
and 8 complex fermions
\be
{\bar\xi}\;,\; {\bar\rho}{}^i_0\; , \; {\bar\zeta}{}^i\;,\;\bar\eta_0 \;.
\ee

Now we turn to solving the on-shell constraints which have the form
\be
\left(D - p^i\psi^i \right)\varphi = 0~, \quad
\left(m^2 - p^2_0 + p^ip^i \right) \varphi = 0~. \label{on}
\ee
Being rewritten in terms of $N=2$ superfields
$\left\{ u,{\bar\rho}{}^i,{\bar v}{}^i,{\bar\eta}\right\}$, they read
\bea
&& Du=\frac{1}{p_0-m}p^i{\bar\rho}{}^i \; , \; D{\bar v}{}^i=\frac{1}{p_0-m}p^i{\bar\eta} \;, \nn
&& D{\bar\rho}{}^i=\frac{1}{p_0-m}\epsilon^{ijk}p^j{\bar v}{}^k\; , \; D\bar\eta =0 \;.
\eea
These conditions put all the fields on the mass shell
\be
(m^2 -p_0^2 +p^i p^i)(\mbox{All bosons})=0 \; , \;
(m^2 -p_0^2 +p^i p^i)(\mbox{All fermions}) =0
\ee
and add the following constraints
\bea
\underline{\mbox{Bosons:}} & & {\bar\phi}{}^i=\frac{1}{p_0-m}\epsilon^{ijk}p^j{\bar v}{}^k_0\;, \;
   {\bar\omega}=0 \; , \nn
\underline{\mbox{Fermions:}} & & \bar\xi = \frac{1}{p_0-m}p^i {\bar\rho}{}^i_0\;, \;
    \bar{\zeta}{}^i=\frac{1}{p_0-m}p^i {\bar\eta}_0 \;. \label{onconstr1}
 \eea
Therefore on shell we have 4 complex bosons $\left\{ u_0\;,\;{\bar v}{}^i_0 \right\}$
and 4 complex fermions $\left\{ {\bar\rho}{}^i_0\; , \; \bar\eta_0 \right\}$.
This is in a nice agreement with the on-shell content found in the
end of the previous Section. To see this in more detail, one should take
into account that, as a consequence of \q{off} and \q{on}, the
longitudinal ($\sim p^i$) parts of the $N=2$ superfields $\bar\rho^i$
and $\bar v^i$ are expressed as spinor derivatives of $u$ and $\bar\eta$.
Then the irreducible set of on-shell $N=2$ superfields at $p^ip^i \neq 0$
is as follows
\bea\label{oncomponents}
&& u=u_0+\theta \frac{1}{p_0-m}(p^i {\bar\rho}{}^i_0) -
\frac{1}{2}\theta\bar\theta (p_0+m)u_0 \; , \nn
&& {\bar\rho}{}^i_{\bot}= {\bar\rho}{}^i_{0 \bot}+
\theta \frac{1}{p_0-m}\epsilon^{ijk}p^j{\bar v}{}^k_{0 \bot}
- {1\over 2}\theta\bar\theta (p_0+m){\bar\rho}{}^i_{0 \bot} \;, \nn
&& {\bar v}{}^i_\bot={\bar v}{}^i_{0 \bot}+
\bar\theta \epsilon^{ijk}p^j{\bar\rho}{}^k_{0 \bot}+
{1\over 2}\theta\bar\theta (p_0+m){\bar v}{}^i_{0 \bot}\;, \nn
&& \bar\eta=\bar\eta_0 - \bar\theta (p^i {\bar v}{}^i_0)+
 {1\over 2} \theta\bar\theta (p_0+m)\bar\eta_0\;.
\eea
One can readily establish the correspondence with the
wave functions \q{a} - \q{c} (up to factors containing $p_0 - m$)
\be
\Phi \sim u_0~, \; \Psi \sim \bar\eta_0~, \;
\Psi^i_\bot \sim \bar{\rho}^i_{0\bot}~, \; \Upsilon \sim (p^i\bar{\rho}^i_0)~,
\; \Phi^i_\bot \sim \bar v^i_{0\bot}~, \; \Xi \sim (p^i\bar v^i_0)~.
\ee
Note that the superfields $\bar{\rho}^i_\bot$ and $\bar{v}^i_\bot$ are not
independent: they describe the same on-shell $N=2$ supermultiplet
$\bar{\rho}^i_{0\bot}(x), \; \bar{v}^i_{0\bot}(x)$ and are related by
$$
\bar{\rho}^i_\bot = \frac{1}{{\vec{p}}{\;}^2}
\bar D\left(\epsilon^{ikl}p^k\bar{v}^l_\bot\right)~, \quad
 \bar{v}^i_\bot = -\frac{(p_0 -m)}{{\vec{p}}{\;}^2}
D\left(\epsilon^{ikl}p^k\bar{\rho}^l_\bot \right)~.
$$

It is worth noting that the superfield wave function
$\varphi (x,\theta, \psi)$ could be chosen fermionic rather than bosonic,
with the corresponding exchange of Grassmann parities between the component
wave functions. This freedom is of the same kind as a freedom of choosing
either $\tau$ or $-\tau$ as the $Z_2$ grading operator in the fixed-gauge
quantization. Also notice that one could put $\varphi$ into some non-trivial
representation of $SO(3)$ by attaching an extra $SO(3)$ index to it.
In this way a reacher $SO(3)$ structure of the final wave functions
can be achieved.

Finally, it is instructive to consider the ``vacuum'' solution within
the GB quantization framework. It is singled out by the additional constraints
$$Q \varphi_{vac}={\bar Q} \varphi_{vac}=0~,$$
where
$$
Q = \frac{\partial}{\partial \theta} + {1\over 2}(p_0 + m)\bar \theta~,
\quad
\bar Q = \frac{\partial}{\partial \bar\theta} +
{1\over 2}(p_0 + m)\theta~, \quad \{Q, \bar Q \} = p_0 + m
$$
(cf. \q{vacstate}). It is straightforward to see that they imply, for
all the component fields, the additional condition
$$(p_0 + m) \left(\mbox{All components} \right) =0~, \; \Rightarrow \;
\Delta \left(\mbox{All components} \right) =0~.
$$
Besides, they require all components in the $\theta,\bar\theta$
expansions in \q{components}, except for the first ones, to vanish.
The latter requirement gives rise to the following relations and vacuum
solutions
\bea
&& (\mbox{a}) \;\left\{
 \begin{array}{l}
 p^i u_0=0 \\
 p^i \bar \eta_0=0
 \end{array}
 \right. \quad
 \Rightarrow \quad
 \left\{
 \begin{array}{l}
u_0 = const\; e^{-imx^0} \\
\bar \eta_0 = const\; e^{-imx^0}
\end{array}
\right. \nn
&& (\mbox{b}) \;\left\{
 \begin{array}{l}
 \epsilon^{ijk}p^j{\bar v}{}^k_0=0, \;p^i\bar{v}^i_0 = 0 \\
 \epsilon^{ijk}p^j{\bar \rho}{}^k_0=0, \; p^i\bar{\rho}^i_0 = 0
 \end{array}
 \right. \quad
 \Rightarrow \quad
 \left\{
 \begin{array}{l}
{\bar v}{}^i_0= e^{-imx^0}p^i{\cal B}(\vec{x}), \;\Delta {\cal B} = 0\\
{\bar \rho}{}^i_0= e^{-imx^0}p^i{\cal F}(\vec{x}), \; \Delta{\cal F} = 0
\end{array}
\right.~. \label{GBvac}
\eea
The solutions (a) correspond to ``trivial'' vacua \q{triv}, while
(b) to the vacua \q{bosVac}, \q{fermVac}.

Let us clarify the precise meaning of the PBGS phases associated
with these vacuum solutions.

Before quantization,
the worldline action \q{action} in a ``static'' gauge $\tau = x^0$ and
with the $\kappa$ symmetry fully fixed by the gauge condition \q{gauge}
(implemented at the classical level), can be considered as the minimal
action of the Goldstone $N=2$ multiplet $x^i(\tau), \psi^i(\tau)$
corresponding to a nonlinear realization of the $d=1$ PBGS option
$N=8 \rightarrow N=2$ \cite{dik}. After quantization of the model
associated with this action we obtained, as the space of
quantum states, the above set of on-shell $N=2$ multiplets which
are combined into a linear on-shell $N=8$ multiplet. Thus, proceeding from
a nonlinear realization of $N=8$ supersymmetry in one dimension, we have
finally arrived at a {\it linear} realization of this supersymmetry
on a set of $N=2$ superfields bearing dependence on all four target
space bosonic coordinates $x^0, x^i$.

An outcome of quantization of the 1/4 superparticle in question
admits the standard interpretation as a first-quantized free
supersymmetric field theory model in $d=4$. The ``1/4 BPS''
conditions \q{GBvac} extract those classical solutions of
the free equations of motion which have a minimal energy and
spontaneously break some of the involved symmetries. After shifting the
superfields by the corresponding condensates, one can expect to find the
relevant Goldstone excitations in the spectrum as collective coordinates
related to the spontaneously broken generators. In particular, for the
condensate (\ref{GBvac}b) one can expect to recover the original worldline
Goldstone multiplet in a new setting, within a {\it linear}
realization of the original 1/4 PBGS option.

To see that this indeed occurs, let us restrict our attention to
the bosonic condensate in (\ref{GBvac}b) (it is unclear how
to interpret the alternative Fermi condensate within the field-theory
framework; normally, the spontaneous supersymmetry breaking is
induced just by bosonic condensates).
We pass to the new ``shifted'' $N=2$ superfield ${\hat{v}}^i(x,\theta)$
\bea
{\hat{v}}^i &\equiv & {\bar v}^i - e^{-m x^0}p^i {\cal B}(\vec{x}) =
{\hat{v}}^i_0 + \theta \frac{1}{p_0 - m} p^i \bar \eta_0 + \bar\theta
\epsilon^{ijk}p^j \bar\rho_0^k \nn
&& + \theta\bar\theta\left({1\over 2}(p_0 +m)
{\hat{v}}^i_0 - \frac{1}{p_0 - m} p^i(p^k{\hat{v}}_0^k)\right)~,
\eea
and observe that under the broken $p^i$ translations (with the parameters
$a^i$) and $Q^i, \bar Q^i$
supertranslations the fields ${\hat{v}}^i_0(x)$ and $\bar\eta_0(x)$ are
transformed as
\bea
\delta_a {\hat{v}}^i_0(x) = ia^k p^k {\hat{v}}^i_0(x) +
ia^kp^k p^i{\cal B}(\vec{x})
e^{-imx^0}~, \quad \delta_\epsilon \bar\eta_0(x) = 2 m e^{-imx^0}
\epsilon^kp^k {\cal B}(\vec{x}) + \ldots ~, \label{inho}
\eea
where dots stand for terms which are linear in fields and vanish under
restriction to the condensate ${\cal B}(\vec{x})$. These transformation
laws directly stem from the $Q^i, \bar Q^i$ transformation of
$\varphi(x,\theta,\psi)$,
$$
\delta_\epsilon \varphi = (\epsilon^iQ_i +  {\bar\epsilon}{}^i\bar Q_i)\varphi~,
$$
$$
Q_i = \frac{\partial}{\partial \psi^i} +{1\over 2}(p_0 - m){\bar\psi}_i +
\theta p_i~, \quad
\bar Q_i = \frac{\partial}{\partial {\bar\psi}{}^i} +
{1\over 2}(p_0 - m){\psi}_i +
\bar\theta p_i~,
$$
rewritten in terms of $N=2$ superfields \q{components} with taking account of
the on-shell relations \q{onconstr1}. The inhomogeneous transformation laws
\q{inho} suggest the following decomposition
\be
{\hat{v}}^i_0(x) = iy^k(x^0)e^{-imx^0}p^kp^i{\cal B} + ...~, \quad
\frac{1}{p_0 - m}\bar\eta_0(x) = -\lambda^k(x^0)e^{-imx^0} p^k {\cal B} + ...~,
\label{decompo}
\ee
with
\be
\delta_a y^i(x^0) = a^i~, \quad \delta_\epsilon \lambda^i(x^0) = \epsilon^i +
...~,  \label{gold1}
\ee
where dots in \q{gold1} stand for terms vanishing upon restriction to
the condensate and those in \q{decompo} for the homogeneously
transforming parts of the fields.
The bosonic and fermionic collective coordinates $y^i(x^0), \; \lambda^i(x^0)$
form a closed multiplet of the unbroken $N=2$ supersymmetry,
\be
\delta y^i = -\epsilon \, \lambda^i~, \quad \delta \lambda^i =
{1\over 2} \bar\epsilon\, p_0\, y^i~.
\ee
It is a linear realization counterpart of the above mentioned
Goldstone multiplet $x^i(\tau), \psi^i(\tau)$ of the original 1/4 PBGS
model.

It should be pointed out that this consideration is purely
kinematical, since we deal with a free $d=4$ superfield theory. In
realistic models of linear realizations of PBGS the vacuum condensate
should arise dynamically as a sort of solitonic solution to self-interacting
theory, with the Laplace equation in \q{GBvac} being replaced by some
nonlinear equation.  In such models, the brane-like Lagrangians of collective
Goldstone modes appear as the leading low-energy approximation of the
full nonlinear Lagrangian  (see, e.g., \cite{towns,ika}). In order to gain
an interacting superfield theory as the result of quantization,
we should start from a generalization of the worldline action \q{action}
containing couplings to an external background and, perhaps, some
potential terms.

\setcounter{equation}{0}
\section{Concluding remarks}
%\vspace{0.4cm}

To summarize, in this paper we examined quantum mechanics of
a massive superparticle model with 1/4 partial breaking of global
supersymmetry which propagates in four--dimensional flat space--time.
The spectrum was shown to contain a finite number of quantum states.
This is in contrast to the massless twistor superparticle example
realizing a 3/4 PBGS option~\cite{bls} where infinitely many (massless)
excitations are known to arise. Although the mass--shell condition is
held in the model, the spectrum resembles very much the non--relativistic
supersymmetric quantum mechanics. In particular, we found a connection
between the states and differential forms on a manifold, similar
to that given in
Ref.~\cite{witten}. This connection implies a geometric interpretation for the
generators of the unbroken supersymmetry as external differentials.
The vacuum states for the case at hand proved to be related to the exact
harmonic one-- and two-forms on the $x$-manifold.
%while the generators of the
%$SO(3)$ rotations turn out to be spontaneously broken.

It is worth noting that all the ingredients of our consideration here,
in particular, the superalgebra \q{algebra}, admit a straightforward
extension to a supersymmetry containing $n+1$ complex supercharges $Q, Q^i$ and
$n+1$ real target bosonic translation generators $P_0, P_i$, $i = 1,\ldots , n$,
with $SO(n)$ being the only space-time symmetry group. In this generic case
we still have one complex $\kappa$-symmetry, and so it corresponds
to the 1/(n+1) PBGS option. Another model which would be of interest
to quantize along the lines of the present paper is the second
$N=8 \rightarrow N=2$ model of Ref. \cite{dik}. As distinct from the
system considered here, this alternative $1/4$ PBGS model does not admit
a straightforward generalization to higher-dimensional supersymmetry.
As a first step, one has to construct the relevant worldline
$\kappa$-invariant action which is still missing.

As for other possible developments, a generalization to
manifolds of nontrivial topology and curved manifolds,
as well as the construction of couplings to external background
(super)fields would be natural next tasks.
%The trick of
%Ref.~\cite{fub} could be relevant here.
A generalization to the branes is also an obvious tempting point.
In particular, there remains the problem of finding out explicit links
with intersecting branes.

\vspace{0.5cm}

\noindent{\Large\bf Acknowledgements}\\

\noindent E.I. thanks A. Pashnev for an enlightening correspondence.
This work was partially supported by the Fondo Affari Internazionali
Convenzione Particellare INFN-JINR,
grants RFBR 99-02-18417,  RFBR-CNRS 98-02-22034, INTAS-00-0254, NATO Grant
PST.CLG  974874  and PICS Project No. 593.

\end{document}